# An Artistic Visualization of Music: Modeling a Synesthetic Experience

Matthew J. Adiletta
Electrical and Computer Eng. and Computer Science
Worcester Polytechnic Institute
mjadiletta@wpi.edu

Oliver Thomas
Mechanical Engineering
Worcester Polytechnic Institute
omthomas@wpi.edu

*Abstract* — This project brings music to sight. Music can be a visual masterpiece. Some people naturally experience a visualization of audio - a condition called synesthesia. The type of synesthesia explored is when sounds create colors in the "mind's eye." Project included interviews with people who experience synesthesia, examination of prior art, and topic research to inform project design. Audio input, digital signal processing (including Fast Fourier Transforms (FFTs)) and data manipulation produce arguments required for our visualization. Arguments are then applied to a physics particle simulator which is re-purposed to model a synesthetic experience. The result of the project is a simulator in MAX 8, which generates a visual performance using particles by varying each particle's position, velocity, and color based on parameters extracted via digital processing of input audio.

## I. INTRODUCTION

This project brings music to life. Since 1877 when Thomas Edison invented the phonograph, music has been about transferring sound from a recorded media, to an audible experience. Modern technology brings music to us over the internet. In fact, over 75% of the music industrys' revenue is made via audio streaming [1]. That said, valuable aspects of experiencing music are lost due to audio only digitization.

Music is an art. Live performances are not just live songs played, but visual experiences. A musician on stage becomes an actor, using his or her instrument to convey emotion. Visual features of performances inform the audience during the concert of music, which are unfortunately not included in the audio digitization of song. This project seeks to create a new and exciting way of visualizing music.

The artistic style used in this project models the human mental experience of synthestegia. Synesthesia is a condition "in which stimulation of one sensory or cognitive pathway leads to involuntary experiences in a second sensory or cognitive pathway" [2]. The specific type of synesthesia we modeled is sound to color synesthesia, where a person hears sound and associates color involuntarily.

Synesthesia is unique to each person, so there is no singular example that encompasses all forms of sound to color synesthesia. Thus, our goal is to find trends between individual synesthetes, and model those trends to create a unique visualization of music. In this report, we present our research and our artistic implementation.

## II. RESEARCH

While researching synesthesia and music to color mappings, we identified a few trends. The trends we focused on are listed below:

1. Emotion to Color
2. Frequency to Color
3. Frequency to Perceptual View
4. Note to Color
5. Dynamics to Color
6. Articulation to Shape

During our research we were able to interview a few synestates about their unique experiences with synesthesia. The interviews were semi-structured, allowing us to ask additional questions if they came up in conversation. We describe our findings below. The questions asked in the interviews are listed in Appendix A.

### A. Interview with Christina Eve

Christina is a synesthete that shares a neural link between her hearing and sight which she described as chromesthesia. She was studying as a musician in college when she realized not everybody shared her unique ability to visualize music through vivid colors and intricate shapes. These visualizations are known as photisms. She is now a visual artist, painting the colors and shapes she sees to share her musical experiences. The purpose of this interview was to ask Christina about the motion she sees in her photisms. We wished to use this along with other information, such as what colors she associates with what sounds, to create a general structure to follow when programming our music visualizer. Please note that everything described below is unique to Christina and her form of chromesthesia; the structure below is not true for every case of synesthesia.

Through the interview, we learned that Christina does not associate individual colors with individual notes. Rather, every song generates a certain theme of colors which typically vary between instruments and their frequencies. These themes are unique between songs, meaning that even music within the same genre may result in an entirely new set of colors. She explained that generally for every song she sees lower tones as darker, more oceanic colors such as blue, green, purple, and black and higher tones as brighter, more metallic colors, common examples being hot pink and silver. She described the deeper tones to be lower in frame and wider in scope and the higher tones to be higher in frame and narrower in scope. She explained how the visualizations appear to be floating in space and can move in all planes (3-Dimensional).

For shapes, she described that percussion is typically very angular and appears as sharp triangles poking out in all different directions. Most instruments, however, appear smooth. One example is an electric guitar which, for her, appears as a snaking line. She explained how circles or bubbles are very common in her visualizations and that typically they converge at one location and then, in her words, "explode and radiate outwards". An example of this is her painting of the band alt-J's "Warm Foothills" [10]. Another typical action in her visualizations is to have a bright light source that ungulates, grows, and or shrinks in between darker colors. An example of this is her painting of Moses Sumney's "Doomed" [10].

An important aspect she described is that these visualizations do not impair her vision whatsoever. She explained how her photisms are in her "mind's eye" and related it to the ability to picture the faces of close relatives or loved ones when they are not present. She also explained how her photisms appear exactly the same every time she listens to a song. The only time a photism may differ for a song is if she is listening on a different sound system. A higher quality sound system may reveal additional instruments and sounds which would add to the visualization and vice versa for a lower quality sound system [11].

*B. Interview with Wendy Case*

Wendy Case is an accomplished musician, specifically a violinist, and like Christina Eve, experiences chromesthesia. Wendy described her interesting case of synesthesia to be developed when learning to read music. Instead of directly associating a note to a color, she associated each note with a specific color and then that color to a sound. This interview's purpose was to gain additional information about a different example of chromesthesia to build upon our structure for the program. Once again, the information below is unique to Wendy Case and should not be generalized with all cases of synesthesia.

Wendy, unlike Christina, correlates individual notes to specific colors. One example she explained is C which appears bluish white for lower C's and yellowish white for higher C's. She stated that snares and other percussion show as black and white since their sounds are not related to a specific note. Wendy also generally described lower frequencies as darker and higher frequencies as brighter. She explained how, within her frame, the lower sounds appear on the left and the higher sounds appear on the right, like a piano or violin. Specifically, when playing chords ascending from low to high, the low notes appear in the bottom left hand corner and the higher notes increase diagonally upwards towards the upper right hand corner.

Wendy explained how she visualizes each note as "pearl-like" flashes. The shape and size of these flashes depends on the amplitude and length of a note. Louder notes appear larger in scale and longer notes result in an elongated flash.

Interestingly, since Wendy learned how to play music by associating the written notes with colors and then those colors to sound, she can also visualize color when reading sheet music. She described that these visualizations appear more vibrant than when she hears music. She also stated that when she listens to music and hears an unfamiliar instrument or note, she does not see flashes for these sounds. Additionally, the more she listens to or reads a musical piece, the duller her photisms become [12].

These two interviews provided a great structure to begin our project. We merged our results and incorporated them into our music visualizer. Utilizing the general themes of lower notes appearing darker and higher notes appearing brighter, notes appearing as pearly or metallic circles, and having these circles explode outwards as they react to the music, our visualization started to look like the descriptions from our interviews. Aspects of synesthesia we could not implement in this version will be incorporated into future versions of this project.

*1. Emotion to Color*

A study from the University of Massachusetts Dartmouth found a relationship between color and emotion. In the study, researchers asked 125 subjects to associate colors that relate to an emotion they feel. They created pie charts showing the percentage of color picked for the respective emotion [3]. Their findings showed the following mappings:

  Anger → Red
  Calmness → Light Blue
  Disgust → Brownish Green / Yellow
  Fear → Black or Red
  Envy → Green
  Happiness → Yellow
  Jealousy → Green
  Sadness → Dark Blue

Another article maps musical keys to emotions. Some interesting mappings are listed below:

  C Major → Innocently Happy
  C Minor → Innocently Sad
  C# Minor → Despair
  D Major → Triumphant
  D Minor → Serious
  D# Minor → Deep Distress
  F Major → Furious
  A Major → Joyful       [4]

With these two studies we see connections between emotion, musical key, and color. We will use this information to inform our decisions with the visualization.

*2. Frequency to Color*

Another trend that we found from descriptions of synthestates are the mappings of frequency to color. Lower frequencies are described as darker, usually consistent with oceanic colors. Higher frequencies are described as brighter with colors such as hot pink, bright yellow, and silver.

These mappings are much more dependent on each individual's perception. These are more general trends we found from analyzing art and reading descriptions of individual cases of synesthesia.

*3. Frequency to Perceptual View*

These mapping are also dependent on the individual's unique synesthesia. A general trend we found is that frequencies are linearly related to perceptual view. This means that if lower frequencies appear at the bottom of a person's perceptual view, then higher frequencies appear at the top of a person's perceptual view. Similarly, lower frequencies may appear on the left side of a person's perceptual view, while higher frequencies may appear on the right side. Generalized directional mappings place the lower frequencies lower and to the left in synesthetes perceptual view, while higher frequencies trend higher and right in synesthetes perceptual view.

This is a very useful characteristic for mapping color to sound for our visualization.

*4. Note to Color*

In an article titled "Color of Sound," mappings of Note to Color are analyzed in detail [5]. Using the Scriabin Correspondence, researchers developed a 1:1 mapping of notes to colors. The general color mappings are shown below:

F → Dark Red
C → Orange Red
G → Orange
D → Yellow
A → Green
E → Light Blue
B → Medium Blue
F# → Royal Blue
Db → Dark Purple
Ab → Light Purple
Eb → Red Violet
Bb → Rosy Brown

Many synesthetes perceive their own 1:1 mapping of notes to colors. This however, is non-algorithmic and dependent on the synethestate.

*5. Dynamics to Color*

This relationship is something that almost every music software incorporates. For example, in MAX MSP, when creating a "~levelmeter" object, MAX generates a level meter with low decibels as green, and higher decibels as red. The red is interpreted as reaching a dangerous level for the listener. This is a simple example of mapping dynamics to color.

From our research, we found synesthetes describe very loud music as "visually active" and in many cases "pearly white" colors. Conversely, soft music is calmer, and more colorful.

*6. Articulation to Shape*

The final relationship we found was articulation to shape. Examples of musical articulation are staccato vs legato. The specific instances of articulation to shape with synesthetes are that concussive bangs are described as "sharper" and "angular" whereas continuous music is described as "flowing." This indicates the following. First, notes that are staccato appear sharp, while notes played legato must appear flowing. Further, if a snare drum is hit, the shape we need to create is angular, sharp, and immediate, whereas, if a violin is playing, the shape we need to create is much smoother and flowing.

### III. IMPLEMENTATION

The implementation of this project is using MAX 8. The reason we chose to use MAX 8 over other programming environments / languages is that MAX is efficient and highly optimized for musical performances, data manipulation, and visualizations. Built in software packages such as jit make visualizing data accessible.

To begin our visualization, we identified aspects of music that we needed to capture in data.

1. Dynamics
2. Frequency Amplitudes
3. Articulation
4. Activity of a Piece of Music

Each of these four types of data can be created through data analysis.

*1. Dynamics*

We captured the dynamics of an audio file using a relatively simple method. We set a minimum and maximum threshold and linearly mapped the current decibel value of the audio to a range between zero and one-hundred to get a "dynamic percentage" of an audio file.

*2. Frequency Amplitudes*

We captured the frequency amplitudes of a song using a Fast-Fourier Transform (FFT). The FFT was challenging to implement. The first question we asked ourselves was, "What range of frequencies should be represented?" Since the human ear can hear from 20 Hz to 20,000 Hz [6], we decided we needed an audio input with a sample rate of at least 40,000 Hz - due to the Nyquist Sampling Rate Theorem [7]. Next, we analyzed the difficulty for identifying individual notes. To do this we would need an FFT with a buffer size that can provide a resolution of 2 Hz, in order to accurately capture each note - a buffer size of at least 20000 values. This is an incredibly large buffer size for the MAX tool. That said, identifying unique notes in a piece of music is a difficult task that requires FFT windowing among other techniques to lead to accurate results. Furthermore, harmonics of notes cause major problems with discriminating the note. Rather than doing a unique note interpretation, we grouped frequencies into 12 unique frequency ranges (four low frequency, four middle frequency, and four high frequency).

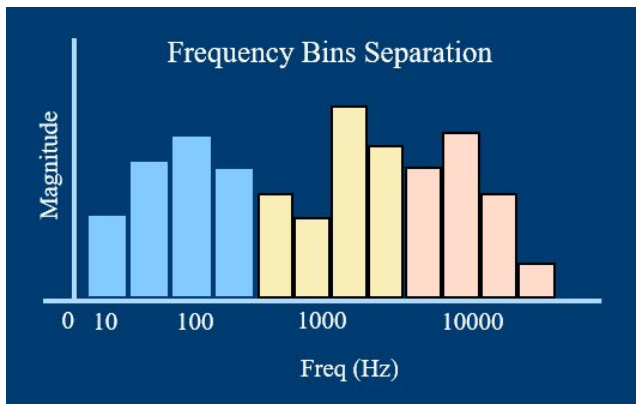

*Figure 1: Visualization of 12 unique frequency bins.*

Each of these frequency groups is called a bin. This gave us the freedom of a smaller buffer size without windowing. Using this approach, we avoided issues with harmonics because our bins captured the major harmonics of the frequency signal.

In order to do the FFT analysis, we used the MAX 8 package jit. Jit has a function jit.FFT which transforms a signal from the time domain into the frequency domain. We first load a buffer of size 1024 with audio input, then compute the FFT using jit.fft. Then we calculate the absolute value of each point in the FFT using cartopol. This converts the real and imaginary parts of the FFT to polar where we have magnitude and angle. For this project, we don't need the angly and only use the magnitude.

Next, we double the first half of the magnitude data points. This is because FFT's only contain 50% unique data. The second half of the data can be "folded" on top of the first half, which is why we double the first half. This produces 512 raw data points we use for the frequency amplitudes.

We further process the FFT amplitudes by combining raw 512 data points into 12 groups. This is accomplished using a cascading splitter (found in the patcher create-bins). We created a cascading splitter using the jit.split function followed by a jit.3m which averages the points. In the first instance of our cascading splitter, we group the first two points of the FFT representing 0 - 86 Hz. Then we average the two points to get our average bin value. Then we scale the bin to normalize across all 12 bins so the "100%" of each bin has the same amplitude value.

Now we have 12 unique frequency bins that we can use for our visualization.

*3. Articulation*

The articulation of a piece of music is hard to capture. One way to think of articulation is how a piece of music changes over time. In the case of a note played with staccato, the derivative of the audio is large, whereas a note played with legato, the derivative of the audio signal is small.

The question becomes, how can we capture the change in an audio signal over time? The approach we took is we first, calculated a running average of the audio frequency bins (p create-avg-bins). Then we took the absolute value of the difference between the current frequency bins (p create-bins) from the averaged frequency bins (p create-avg-bins) which produced articulation data (p current-volatility). In the MAX patcher, this is called the volatility of a signal.

We then computed a running average of the current volatility (p create-avg-volatility-bins). From this patcher, we could see the average volatility of each bin over a certain number of datapoints. We then computed the average volatility over eight data points, essentially deriving the average volatility of each bin over .186 seconds:
 (given: sample rate 44100; buffer size 1024)
 time =  number_data_points * (buffer_size / sample_rate)

We can analyze the average volatility to indicate articulation. If a bin has a high average volatility, then the notes being played are likely staccato / percussive. If a bin has a low average volatility, then the notes being played are likely legato / flowing.

*4. Activity of a Piece of Music*

We define the activity of a piece of music by observing major changes in a piece. For example, if all of a sudden there is a large change in a piece, our visualization needs to react accordingly. Capturing this data is simple based on our previous development. We use the current-volatility patcher and set a threshold. If the current-volatility exceeds a certain threshold such as 70% of maximum, then we trigger an event (p set-volatility-triggers).

*A. Data Visualization*

Now that we have all this data readily available for use, we can create the visualization. The first visualization we created was graphing all the data at once. The visualization shows four graphs: current-frequency-values, averaged-frequency-values, current-volatility-values, averaged-volatility-values. There is also a matrixctl object that indicates which triggers are active (p Visualizing).

*B. Particle Simulation Design*

The 3D visualization technique which provides the best visualization of our parameters is a particle simulation. The simulator defines that each particle is associated with a certain gravity point. Each particle has 3 components: a current 3D position vector, a current 3D velocity vector, and a current RGBA color vector. In our design, the A component of the RGB is not used for color, so we used the A value to associate a particle with a gravity point.

Since there are 12 frequency bins, we created 12 gravity points, separated along the y axis so that the lowest frequency bin has its gravity point moving around y = -12, and the highest frequency bin has its gravity point moving around y = 10.

Each of the 12 particle groups has six inputs:

- uColorRGB
    The RGB value for setting the color of the particle. This is derived using the average-frequency-bin values.
- uTarget
    The (x, y, z) location of the gravity point that the particle is drawn towards. The gravity point is constricted to a 3x3x3 (x, y, z) point space (p target).
- uForceAmt
    The force with which the particle is attracted towards the gravity point. We use this point to accelerate the particle towards the gravity point.
- uColorMag
    This parameter defines the scalar multiplier of the color magnitudes between 0 and 1. The number is derived from the average-frequency-bin values. As an example, if we decrease the number of averaged points, the particle's color becomes more sensitive to the music.
- uEmphasis
    This parameter defines any major changes of the particles associated bin. It is connected to the triggered volatility value. If the trigger is active, then an emphasis occurs which doubles the uForceAmt.
- uYCenter
    This parameter defines the y location of the center of the particle grouping. The range for the particles goes from -12 to 10 in the current model, but it can technically be spread out much wider.

All of these parameters are set in a unique patcher for each frequency bin (pmyShader[Low/Med/High][1/2/3/4]).

There is one final parameter that is passed to the particle simulator that is uniform across all shaders - the uColorSensitivity. This value is a multiplier on the RGB value. It's a nice parameter to fine tune for each individual song. All of these parameters are passed to the file GPU_particles.jxs which defines the particle simulation. In total there are 3 + 12*6 + 1 parameters input to this file:

- iPosition, iVelocity, iColor

- uTarget, uColorRGB, uYCenter, uEmphasis, uColorMag, uForceAmt

- uColorSensativity

The reason that iPosition, iVelocity, and iColor are separate from the other parameters is because jit has a function called jig.gl.tf (transform feedback) which takes Position, Velocity, and Color vectors and feeds the values back on itself each iteration. Essentially, it keeps track of the Position, Velocity, and Color values for the particle at all times. We adjust these values using the six unique parameters in each of the shaders.

*C. Setting Up Run Environment*

Before we run the simulation, we need to set up our environment. First, we need to download the test files. The test files are .aif files that we can use in the simulation which can be downloaded from the git hub: https://arcgit.wpi.edu/case/musicvisualizer. Copy the audio files to the same folder as the particle-simulator.pat.

Next, we need to update our jit version to gl3. Open the package manager and search for gl3. Download the package. Then open Options → Preferences. Scroll to Jitter and change OpenGL Engine from gl2 to gl3. Then restart MAX.

*D. Running the Simulation*

In order to run the simulation, we first need to initialize all the default settings. The important default settings are listed below:

    num-points-to-average = 4
    num-points-to-average-vol = 8
    trigger-val = 70
    max-average = 0.3
    max-trigger = 0.15
    color-sensitivity = 2
    range_max = 0.3

To set / reset these values, press the "bang defaults" button at the top of the patch.

Next we need to render the jit simulation window. Do this by scrolling right to view the largest blue box. Press anim_reset to reset the camera view in the jit window. Activate the toggle to start jit.world.

Next we need to start the particle simulator. Do this by first pressing the button connected to the play audio toggle. This bangs the components which starts the jig.gl.mesh simulator. Then press the space-bar to activate.

Finally, we need to pick a song to play. First press a button corresponding to the song you wish to load. Then toggle the play audio on and off to load the song. Make sure the speaker is active or else the song will not play.

Hopefully the song started and you see activity in the jit window. Click the jit window to allow interaction with the camera. Use "A" "S" "D" "W" "Q" and "Z" to rotate your view in the jit window. The useful motions to get started are pressing and holding "S" to zoom out. Hopefully the song is clear. Then press and hold "Z" to look at the visualization from a "top-down" view.

If ever you feel "lost" in the simulation, press the anim_reset located in the largest blue box of the patcher.

## IV. CONCLUSION

### A. Results

The result of this project is that a user has the ability to render a visual performance of any audio input. The user has the ability to interact with the program through a number of variables that change the look and feel of the performance. The artistic style of the visualization, models that of descriptions from synesthetes, allowing non-synesthetes to see something similar to what they see.

Figure 1 below shows a screenshot of the simulation in action.

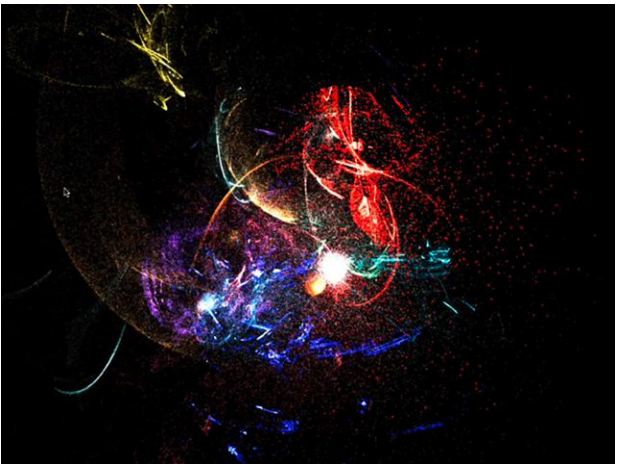

*Figure 2: A screenshot of the particle simulation from a top-down view.*

Some of the tunable parameters the program interfaces with are listed below:

1. The number of averaged points for the fft values for changing the emphasis color and brightness of the particles.
2. The number of averaged points for the volatility values for increasing the velocity of the gravity points.
3. The trigger value for adding emphasis on major events.
4. The max-average value and max-trigger to adjust the program to for major changes in input files - if the max decibel value for one program is significantly different than another file, vary this parameter to map appropriate maximum values.
5. The color sensitivity value which is a scalar multiplier for the brightness of the particles.
6. The base color of the particles in each bin.
7. The audio input.

### B. Future Work

This project provides a great foundation for future work. Future work includes:

1. Add a windowing function to the jit.fft so that the cutoff over time is less dramatic and we see more accuracy with the FFT.
2. Deeper connection between the research on synesthesia and the associated colors of certain songs during a performance.
3. Create an interface that can vary the tunable parameters for performances more easily.
4. Create presets for certain songs categories, moods and styles.
5. Implement a new system for how the colors change with changes in frequency values. Currently only one RGB value is allowed to change with the frequency. Code a patch that allows more than one of the RGB values to change.
6. Create a format of the visualization in augmented reality.

APPENDIX A

1. Do you associate specific notes with colors?
    a. If so, could you describe what colors correspond with what notes?
2. Do the colors you associate with music change based on the theme/emotion of the music?
    a. If so, are these colors constant? Or do they vary between songs?
    b. Do the colors you associate with the themes of music correlate with what people generally connect with emotion? I.e. blue with sadness or red with anger?
3. Are there specific shapes/formations that you correlate with specific sounds?
    a. Do shapes/formations always take the same color or does this vary?
4. Do the shapes and colors you visualize appear dynamic with fluid motions? Or do they appear more as flashes? Or does this vary between song and sound?
5. Have you noticed a certain trend of shape, color, or motion that occurs with varying frequencies? I.e. higher or lower tones?
6. Does the amplitude of music affect the color or shape of the visualizations?
7. Is there a specific genre of music that produces better visualizations? Are there genres that produce more dull and faded visualizations? Or don't produce visualizations at all?
8. When listening to a song, is the visualization always the same, or does it vary a bit or even a lot between listens?
9. Are the visualizations in 3 Dimensions? Or do they remain in 2 Dimensions?